\documentclass[12pt,preprint]{aastex}

\newcommand{\keV}{\mbox{ keV}}
\newcommand{\kms}{\mbox{ km s}^{-1}}
\newcommand{\col}{\mbox{ cm}^{-2}}
\newcommand{\La}{L$\alpha$}  
\newcommand{\Ka}{K$\alpha$}  
\newcommand{\Ha}{H$\alpha$}  
\newcommand{\pcc}{\mbox{ cm}^{-3}}
\newcommand{\NLa}{\ion{N}{7} L$\alpha$}
\newcommand{\OKa}{\ion{O}{7} K$\alpha$}
\newcommand{\OLa}{\ion{O}{8} L$\alpha$}
\newcommand{\NeKa}{\ion{Ne}{9} K$\alpha$}
\newcommand{\NeLa}{\ion{Ne}{10} L$\alpha$}
\newcommand{\MgKa}{\ion{Mg}{11} K$\alpha$}
\newcommand{\MgLa}{\ion{Mg}{12} L$\alpha$}
\newcommand{\SiKa}{\ion{Si}{13} K$\alpha$}
\newcommand{\SiLa}{\ion{Si}{14} L$\alpha$}
\newcommand{\HII}{\ion{H}{2}}

\begin{document}

\title{The X-ray Spectrum of Supernova Remnant 1987A}

\author{Eli Michael\altaffilmark{1}, Svetozar
Zhekov\altaffilmark{1,2}, Richard McCray\altaffilmark{1}, Una
Hwang\altaffilmark{3}, David N. Burrows\altaffilmark{4}, Sangwook
Park\altaffilmark{4}, Gordon P. Garmire\altaffilmark{4}, Stephen
S. Holt\altaffilmark{5}, and G\"{u}nther Hasinger\altaffilmark{6}}

\altaffiltext{1}{JILA, University of Colorado, Campus Box 440,
Boulder, CO 80309-0440; michaele@colorado.edu.} 

\altaffiltext{2}{On leave from Space Research Institute, Sofia,
Bulgaria.}

\altaffiltext{3}{NASA Goddard Space Flight Center, Code 662,
Greenbelt, MD 20771.}

\altaffiltext{4}{Department of Astronomy and Astrophysics,
Pennsylvania State University, 525 Davey Laboratory, University Park,
PA 16802.}

\altaffiltext{5}{F. W. Olin College of Engineering, Needham, MA 02492}

\altaffiltext{6}{Max-Planck-Institut f\"{u}r
Extraterrestrische Physik, Postfach 1312, D-85748 Garching, Germany.}

\begin{abstract}

We discuss the X-ray emission observed from Supernova Remnant 1987A
with the {\it Chandra X-ray Observatory}.  We analyze a high
resolution spectrum obtained in 1999 October with the high energy
transmission grating (HETG).  From this spectrum we measure the
strengths and an average profile of the observed X-ray lines.  We also
analyze a high signal-to-noise ratio CCD spectrum obtained in 2000
December.  The good statistics ($\approx 9250$ counts) of this
spectrum and the high spatial resolution provided by the telescope
allow us to perform spectroscopic analyses of different regions of the
remnant.

We discuss the relevant shock physics that can explain the observed
X-ray emission.  The X-ray spectra are well fit by plane parallel
shock models with post-shock electron temperatures of $\approx 2.6
\keV$ and ionization ages of $\approx 6 \times 10^{10}$ cm$^{-3}$
s.  The combined X-ray line profile has a FWHM of $\approx 5000
\kms$, indicating a blast wave speed of $\approx 3500 \kms$. At this
speed, plasma with a mean post-shock temperature of $\approx 17 \keV$
is produced.  This is direct evidence for incomplete electron-ion
temperature equilibration behind the shock. Assuming this shock
temperature, we constrain the amount of collisionless electron heating
at the shock front at $T_{e0} / T_s = 0.11^{+0.02}_{-0.01}$.  We find that the
plasma has low metallicity (abundances are $\approx 0.1$ -- $0.4$
solar) and is nitrogen enriched (N/O $\approx 0.8$ by number), similar
to abundances found for the equatorial ring.

Analysis of the spectra from different regions of the remnant reveals
slight differences in the parameters of the emitting plasma.  The
plasma is cooler near the optical Spot 1 (at position angle $\approx
30^{\circ}$) and in the eastern half of the remnant, where the bright
optical spots are found, than in the western half, consistent with the
presence of slower ($\approx 500 \kms$) shocks entering denser ring
material. There is an overall flux asymmetry between the two halves,
with the eastern half being 15 -- 50\% brighter (depending on how the
center of the remnant is defined).  However, our spectroscopic
analysis shows that $<5\%$ of the overall X-ray emission
could come from a slow shock component.  Therefore the flux asymmetry
cannot fully be due to X-rays produced by the blast wave entering the
ring, but rather indicates an asymmetry in the global interaction with
the circumstellar material interior to the ring.

\end{abstract}


\section{Introduction}

Supernova (SN) 1987A provided the astrophysical community with a
wealth of information about core collapse supernova (see
e.g., \citealt{arne89,mccr93}).  Now, as the object enters its remnant
phase, it will reveal still more about the physics of the explosion
and of the progenitor star's history, as well as give insight into the
processes occurring in other supernova remnants (SNR).  The radiation
from SNR 1987A is now dominated by the interaction of the SN debris
with the circumstellar material (CSM) which was present before the
star exploded.  This interaction will convert the kinetic energy
stored in the debris expansion into radiation which is observable from
the radio to X-ray bands.

The circumstellar gas is most obvious in the triple ring system seen
in optical wavelengths \citep{burr95} as it recombines following the
initial ionizing flash from the SN \citep{lund91}.  The mechanism responsible
for forming these rings is still uncertain. In one model the system is
a result of the self interaction of the progenitor star's stellar winds
\citep{luo91,wang92,blon93}. The appearance of both radio
\citep{stav92,stav93} and X-ray \citep{beue94,gore94,hasi96} emission
from the remnant at day $\approx 1000$ suggests that the supernova
blast wave encountered an \HII\ region interior to the equatorial ring
\citep{chev95,bork97a}.  This \HII\ region was probably formed by the
evaporation of gas off of the ring by the blue supergiant progenitor
star \citep{chev95}.

\begin{figure}
\epsscale{.5}
\plotone{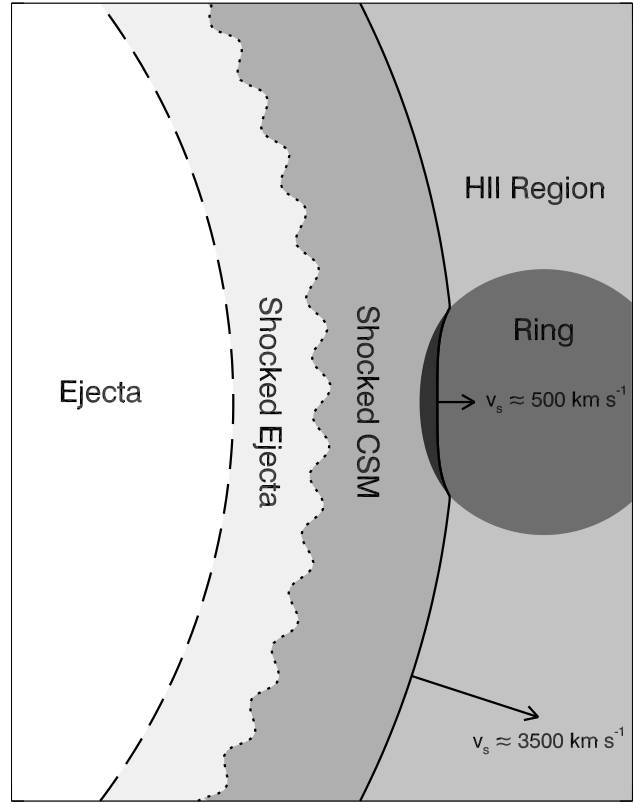}
\caption{A schematic of SNR 1987A's structure showing a cross section 
perpendicular to the equatorial (ring) plane.  The grayscale depicts density,
with darker shades showing denser material.  The blast wave (solid line) 
travels into the CSM, while the reverse shock (dashed line) travels into
the SN ejecta.  The regions of shocked CSM and shocked ejecta are
separated by an unstable contact surface (dotted line).  Traveling through 
the  \HII\ region the blast wave has a velocity of $\approx 3500 \kms$, but
where it encounters denser ring material it slows to $\approx 500 \kms$.
\label{fig-DSS}}
\end{figure}

The hydrodynamics of the interaction is quite complex
\citep{chev92,bork97b} and depends on the density distributions of
the expanding debris and the CSM.  In any case, a double shock
structure forms which consists of a blast wave which travels
outwards into the CSM and a reverse shock which travels back (in a
Lagrangian sense) into the SN ejecta (see e.g., \citealt{chev82}).  This
structure is shown schematically in Figure~\ref{fig-DSS}.  Between
these two shocks is a region of shocked circumstellar gas and a region
of shocked ejecta gas, separated by a contact surface which is
hydrodynamically unstable \citep{chev92}.

Emission from the reverse shock has been observed as high
velocity ($\sim \pm 12,000 \kms$) \La\ and \Ha\ emission
\citep{mich98} created as neutral hydrogen crosses the shock front.
By modeling the line profiles \citet{mich98} found that emission from
the reverse shock front is confined primarily to the equatorial plane
($\pm 30^{\circ}$) and is located at $\approx 75\%$ of the distance
from the SN to the inner surface of the equatorial ring.  Images of
SNR 1987A in radio bands \citep{gaen97,manc01} show a toroidal
geometry with an east-west brightness asymmetry.  These images support
the notion that the emission from the remnant is brightest in the
equatorial regions.  \citet{manc01} also find that the radio image is
expanding with a radial velocity of $\approx 3500 \kms$.  This radio
emission is most likely synchrotron emission from nonthermal
particles which are accelerated in the shocks.  While it is uncertain
where this emission originates, it is reasonable to assume that it
comes from the approximate region between the blast wave and
reverse shock.  If so, the expansion speed of the radio remnant is a
good approximation for the speed of the blast wave in the \HII\
region.

Hydrodynamic models for the interaction of the SN 1987A's blast wave
with an equatorial \HII\ region produce blast wave velocities of
$\approx 4100 \kms$ \citep{bork97a}, close to that inferred from the
radio images.  Such a shock can produce X-ray emitting plasma with
electron temperatures of several keV.  X-ray images taken by the {\it
Chandra X-ray Observatory} ({\it Chandra}) were shell-like and
confirmed that the emission was indeed from the shocks rather than
emission from a central source \citep{burr00,park01}.  The X-ray
spectrum is characterized by electron temperatures of $\approx 3
\keV$. \citet{park01} showed that the hard band X-ray image ($E = 1.2$
-- $8.0 \keV$) of the remnant looks very similar to the radio image,
which suggests that the double shock structure is the source of
emission in both bands.  X-ray images from {\it Chandra} also suggested an
expansion rate of $5200 \pm 2100 \kms$ \citep{park01}, consistent with
the radio image expansion rate and the theoretical models.

One of the young remnant's most exciting developments in the last
several years has been the appearance in the optical and ultraviolet
(UV) of several rapidly brightening unresolved spots on the equatorial
ring \citep{lawr00,mich00}.  These spots appear where the blast wave
has encountered dense ($n > 10^4 \pcc$) obstacles on the inside of the
optical ring.  In these obstacles the blast wave has slowed
considerably (to $\sim 250 \kms$) and cooled sufficiently to emit
copious UV and optical radiation \citep{pun02}.  These radiative
shocks are too slow to produce substantial X-ray emission. There may,
however, be lower density obstacles near the optical spots into which
shocks with speeds of $\approx 500$ -- $1000 \kms$ are transmitted.
Such shocks would be copious producers of soft X-ray emission ($E
\lesssim 1 \keV$) but they would not produce significant amounts of
optical or UV line emission.  The soft band X-ray image ($E = 0.3$ --
$0.8 \keV$) presented by \citet{park01} shows X-ray brightening at
approximately the same locations as the optical spots, suggesting that
intermediate velocity shocks are present in their vicinity.

Here we present and analyze in more detail two X-ray data sets
obtained with {\it Chandra}: a high spectral resolution dispersed spectrum
(obtained with the HETG grating) first shown in \citet{burr00}; and
spatially resolved CCD spectra obtained from the high signal-to-noise
ratio (S/N) imaging observation presented by \citet{park01}. We
present the observations and reductions in \S\ref{sec-obs}. In
\S\ref{sec-physics} we present our shock models to describe the
observed X-ray emission. In \S\ref{sec-analCCD} we analyze the CCD
spectrum to determine the overall shock parameters and their
variations across the remnant.  We also consider spectroscopic
evidence for the existence of slow X-ray emitting shocks. In
\S\ref{sec-analdisp} we present the X-ray line fluxes obtained from
the dispersed spectrum. These data confirm the results of the CCD
spectral analysis and the legitimacy of the shock model for the
emission.  In \S\ref{sec-lineprof} we present the combined X-ray line
profile and compare it to simulated profiles in order to establish the
speed of the blast wave. In \S\ref{sec-disc} we discuss our findings
and what they tell us about the remnant's structure and development,
as well as what we may learn from future investigations. We conclude
in \S\ref{sec-summary} with a summary of our results.

\section{Observations and Data Reduction} \label{sec-obs}

\subsection{Dispersed Spectrum}

As a combined effort of 2 Guaranteed Time Observation programs, 120
ks of telescope time was allocated to obtain a high resolution
dispersed X-ray spectrum of SNR 1987A with {\it Chandra}. On 1999 October 6
two observations (Obsid 124 \& 1387) were made with the HETG grating
and the ACIS-S detector ($E / \Delta E \approx 500$ at 1 keV) with a
net exposure time of $116.1$ ks. Here we work with the reprocessed (2000
September 1) data set.  We screened photon events on energy (0.3 --
8 keV), grade (02346), and status (0). The lightcurve is free of
background flares. We removed spurious `streaking' on the S4 chip,
greatly increasing the S/N of the spectrum while eliminating only a
negligible amount of observing time ($<~0.05$\%). To further increase
the S/N of the spectrum, we removed a region of high detector
background on chip S1, affecting one arm of the MEG spectrum in the
energy range $E \approx 0.52$ -- $0.56 \keV$.  We added this region as
well as additional bad columns and pixels removed by visual inspection
of the data to the bad pixel file in order to properly account for
their removal when creating effective area distributions (ARFs).

We established the position of the source from the centroid of the
0$^{th}$ order image. We resolved spectral orders and assigned
wavelengths to photons in the spectra making use of the {\it CIAO
v2.0} procedures {\it tg\_create\_mask} and {\it tg\_resolve\_events}.
The source is faint enough that only 1$^{st}$ orders of the HEG and
MEG spectra have detectable signal. Based on the cross-dispersion
width of the observed spectra we chose to extract the spectrum from
regions along the dispersion directions $3\arcsec$ wide centered on
the 0$^{th}$ order image.  We use no aperture correction since the
data do not show a statistically significant increase in source counts
with a wider aperture.  Choosing a wider aperture would yield an
extracted spectrum with substantially decreased S/N owing to the
required subtraction of a larger background emission component.

Because SNR 1987A is so faint in the X-ray band, we need to subtract
the background signal carefully. The background consists of a diffuse
background as well as structured background from other faint sources
which overlap with the dispersed spectra. To characterize the local
nature of this background, we use regions along the dispersion
directions extending from $5\arcsec$ to $20\arcsec$ above and below
the source regions.  We chose these regions to be wide enough to
contain enough counts to statistically represent the background, but
narrow enough to represent its local structure. We combined and
extracted the photons in these regions the same way as for the source
spectrum.

\begin{figure}
\epsscale{1.}
\plotone{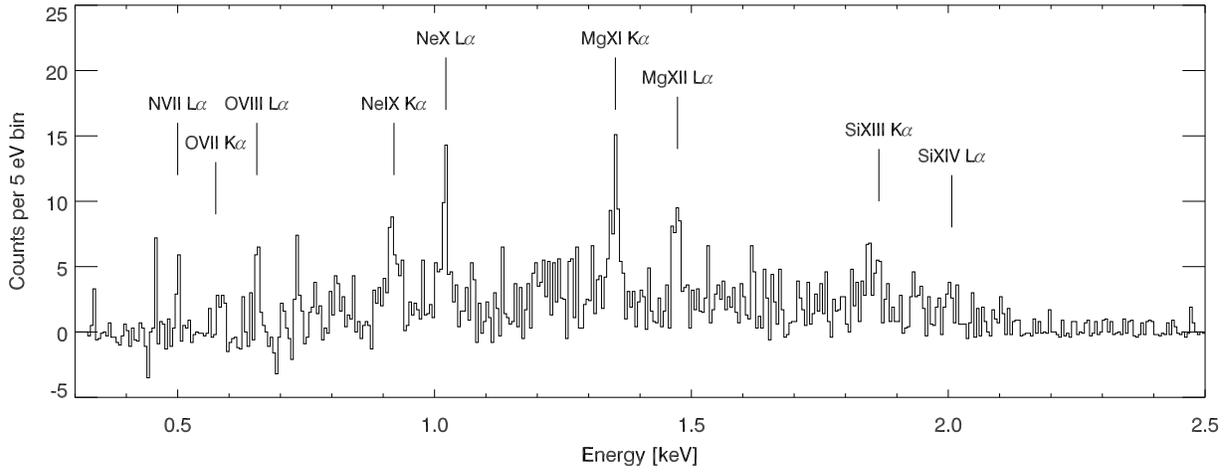}
\caption{Background subtracted dispersed spectrum of SNR 1987A.
Several line identifications are shown. \label{fig-dispspec}}
\end{figure}

To create effective area distributions for both the HEG and MEG
spectra we used the {\it CIAO} script {\it mkgarf\_acis}, the updated
bad pixel file, and the standard response matrices for the ACIS HETG
spectra.  Since both observations were performed with the same roll
angle and pointing they have identical backgrounds and ARFs. Therefore
we have combined the extracted spectra from both observations. We also
combined positive and negative spectral orders (as well as their ARFs)
for both the HEG and MEG spectra. We present the resulting fluxes in
Table~\ref{tab-dispfluxes} and show the extracted spectrum in
Figure~\ref{fig-dispspec}.  We display the combined (MEG and HEG $\pm
1^{st}$ orders) background subtracted count spectrum. While noisy,
lines from H- and He-like ions of N, O, Ne, Mg and Si are easily
identified.

\subsection{CCD Spectra} \label{sec-obsCCD}

On 2000 December 7 we obtained an observation of 99 ks (Obsid 1967)
intended to provide a high spatial resolution, high S/N image of the
X-ray remnant \citep{park01}.  The image was obtained on the ACIS S3
chip ($E / \Delta E \approx 8$ at 1 keV), without a spectral grating
in the optical path.  We processed this data set as in \citet{park01} to
correct for the effects of charge transfer inefficiencies (CTI)
\citep{town00,town01a} and to improve spatial resolution using a
sub-pixelization technique \citep{tsun01}. We screened events on
energy (0.3 -- 8 keV), grade (02346), and status (0). The lightcurve
is free of background flares. The pileup fraction is small ($\lesssim
4$\%) and therefore may be ignored.

\begin{figure}
\epsscale{.5}
\plotone{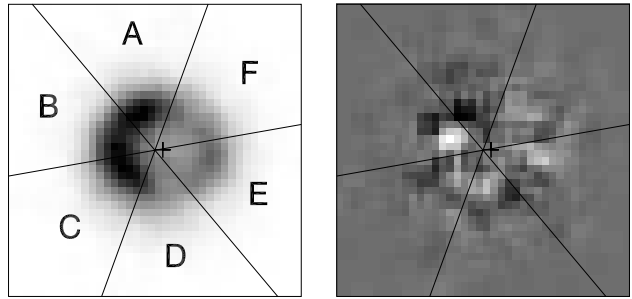}
\caption{The left panel shows the total X-ray count image (N is up).
The right panel shows a `hardness' image created by subtracting the
0.3 -- $0.8 \keV$ image from the 1.3 -- $8.0 \keV$ image (lighter
shades show harder emission).  Both images have 0\farcs1 pixels and
are smoothed by a $3 \times 3$ pixel boxcar.  Overlaid on these images
are the boundaries between the 6 extraction regions (A -- F) used in
our spatially resolved spectroscopic analysis. The regions originate
from the centroid of the image. The pluses mark the center of the
remnant obtained by fitting an ellipse to the image
shape.\label{fig-regions}}
\end{figure}

Figure~\ref{fig-regions} shows the X-ray image which contains $\approx
9250$ source counts.  It confirms the findings of \citet{park01}: the
emission is annular, slightly elliptical, and is brighter in the east
with enhancements at the approximate locations of optical Spots 1 -- 5
\citep{lawr00}.  Also shown in Figure~\ref{fig-regions} is a
`hardness' image created by subtracting the 0.3 -- 0.8 keV image from
the 1.3 -- 8.0 keV image.

Since absolute astrometric registration of the X-ray image with
optical or radio counterparts has not been established at a level
$\lesssim 0\farcs3$, the exact location of the X-ray remnant's center
(i.e. the location of the SN) cannot be determined.  We locate the
center of the X-ray remnant in one of two ways.  The best defined
center is the centroid of the photon arrival locations.  However,
since the emission is highly asymmetric, there is no reason to expect
this point to be coincident with the SN.  Alternatively, we can locate
the center by fitting an ellipse to the shape of the image. The center
of this ellipse lies $\approx 0\farcs2$ west of the image centroid
(see Fig.~\ref{fig-regions}).  In our analysis (\S\ref{sec-analCCD})
we use the image centroid since it is unambiguously defined.
Fortunately, our spectroscopic results do not depend on which center
is used.  However, as we discuss in \S\ref{sec-disc}, the overall
level of flux asymmetry between the western and eastern halves of the
remnant depends greatly on the choice of center.

With an angular resolution of $\approx 0\farcs5$, the telescope allows
us to distinguish the spectra of different regions of the remnant.  In
our preliminary analysis we tried many different apertures. Here we
discuss only three sets of apertures.  First, we take an aperture
which contains the entire remnant. Second, we divide the remnant into
eastern and western halves to investigate the clear asymmetry which is
seen in the image.  Finally, we split the remnant into the 6 regions
shown in Figure~\ref{fig-regions}.  We choose the boundaries between
these regions to match features seen in the `hardness' image.  The
X-ray emission at the location of optical Spot 1 (at position angle
$\approx 30^\circ$) is much softer than that found in the rest of the
remnant.  We thus place region A in order to isolate the emission
coming from the vicinity of Spot 1. Each of the 6 regions in
Figure~\ref{fig-regions} contain between 1300 and 1700 counts, giving
spectra with statistically acceptable numbers of counts.

We determine the background spectrum from a region defined by a
circular annulus with inner radius of $30\arcsec$ and outer radius of
$100\arcsec$ that is free of other sources.  The background is small,
with $<2\%$ of the counts in the source spectrum. We extracted the
spectra using the {\it CIAO} command {\it dmextract}, created an ARF
for the source using {\it asp\_apply\_sim, asphist} and {\it mkarf},
and utilized the response matrix which was created to properly account
for the CTI correction \citep{town01b}.

\section{Physical Picture} \label{sec-physics}

Figure \ref{fig-DSS} depicts a schematic of the remnant's shock
structures.  In reality, the blast wave will vary in speed as it
interacts with density gradients in the CSM, and reflected shocks will
form where the gradients are steep enough. Therefore the hot plasma
behind the shock will not be produced by a single steady-state shock
wave.  However, in the absence of detailed information about this
density distribution, we cannot reasonably attempt to determine the
full distribution of plasma as a function of temperature and
ionization history in the remnant.  Instead, we assume that plane
parallel shock models will be an adequate first approximation to
describe the emission from the remnant.  Since the shell of shocked
gas is expected to be thin ($\Delta R / R \approx 0.1$) relative to
the radius of the remnant the plane parallel approximation may be
acceptable.  We expect that the X-ray emission from the remnant will
be dominated by shocked CSM rather than by shocked ejecta, since
hydrodynamical models of the interaction \citep{bork97a} show that the
emission measure behind the blast wave is a factor of $\sim 5$ larger
than that behind the reverse shock. Then, by fitting the data with
single shock models, we attempt to determine the average plasma
parameters behind the blast wave.

We also consider the hypothesis that the blast wave is interacting at
some locations with denser gas associated with the equatorial ring.
In this case a distribution of shocks with varying speeds and ages
will be present.  We represent this case with a simplified model
consisting of two shock components, one representing the blast wave in
the \HII\ region and the other representing a slower shock transmitted
into the ring.

The passing shock rapidly heats and compresses the pre-shock gas,
causing atoms to be ionized and excited and to emit radiation. As gas
moves downstream from the shock front it will undergo further
ionization until it finally finds itself in ionization
equilibrium. Given enough time, the gas may radiate away a substantial
portion of its thermal energy and become a radiative shock.  However,
the shocks that produce the X-rays from SNR 1987A have not been
present long enough for this to occur, and therefore will be
non-radiative (or adiabatic) shocks. A simple non-equilibrium
ionization (NEI) model will not accurately represent the emission from
the shocked plasma because some gas will have passed through the shock
only recently, while other gas will have passed through long before.
Therefore, the emission comes from a collection of slabs of shocked
gas each with their own NEI age.  The ionization age ($n_e t$) of the
shock as a whole, which is the ionization age of the oldest parcel of
shocked gas, is a crucial parameter in determining the emission from a
shock.  Shocks with lower values of $n_et$ are not able to produce
plasma with higher ionization ages.

In addition to the ionization age, the temperature of the shocked
plasma also affects the emission from the shock.  For adiabatic shocks
the shock temperature is given by $T_s = 3 \bar{m} v_s^2 / 16 k$,
where $\bar{m}$ is the mean mass per particle ($\bar{m} \approx 0.7
m_p$ for fully ionized gas with abundances representative of the
ring). The shocks in SNR 1987A are most likely collisionless shocks,
where thermalization of the fluid's bulk flow is mediated by turbulent
magnetic fields.  In such a shock each element will have a temperature
given by $T_i = 3 m_i v_s^2 / 16 k$.  The electron temperature ($T_e$)
can be orders of magnitude below the ion temperatures. While we assume
that each species will thermalize its bulk motion (such that the
thermal velocity distributions are Maxwellian), it is unclear whether
different species will efficiently share thermal energy at the shock.
Collisionless processes may be present which can partially heat the
electrons at the shock front (e.g., \citealt{carg88}). However, based
on observations of other fast shocks \citep{ghav01}, we expect that
the electron temperature immediately behind the shock front will be
less than the shock temperature ($\beta = T_{e0} / T_s <
1$). Eventually Coulomb collisions in the post-shock flow will cause
all species to equilibrate their temperatures. It is important to note
that the emitted radiation will have an excitation temperature that is
characterized by the electron temperature, not the ion or shock
temperatures.  If Coulomb collisions can cause significant heating in
the post-shock plasma, the emission from a shock must be calculated by
considering both the ionization and temperature history of the
post-shock gas.

The X-ray spectral fitting package {\it XSPEC v11} contains models
which calculate the emission from such shocks \citep{bork01}.  The
{\it pshock} models calculate non-equilibrium ionization and emission
from a shock with constant post-shock electron temperature. The {\it
npshock} models allow unequal electron and ion temperatures at the
shock front, then calculate electron heating through Coulomb
collisions in order to determine the proper distribution of post-shock
electron temperature with which to compute the non-equilibrium
ionization.  We use these two shock models in our analysis of the
observed X-ray emission.

\begin{figure}
\epsscale{.5}
\plotone{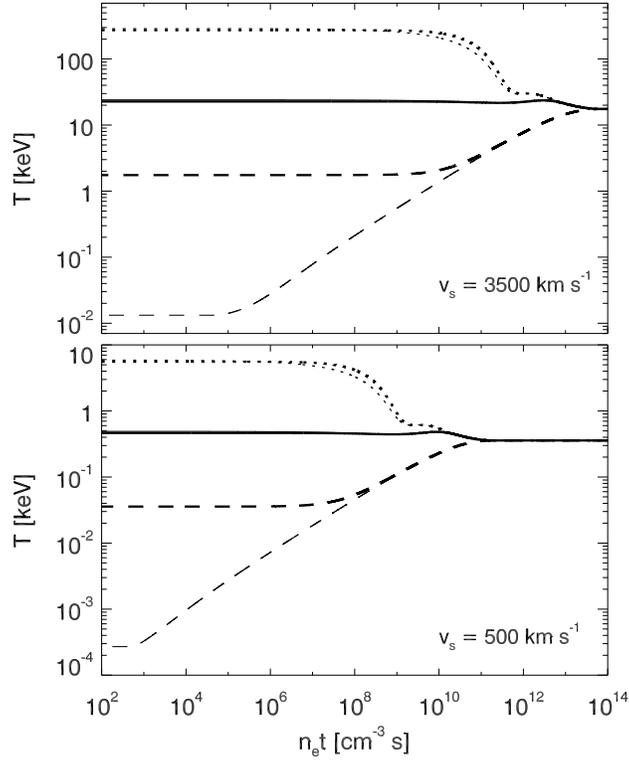}
\caption{Curves show electron (dashed), proton (solid), and Mg
(dotted) temperature equilibration through Coulomb collisions behind
the shock for two different shock velocities.  Thick lines show the
case where partial electron heating occurs at the shock front ($\beta
= 0.1$), thin lines have no electron heating at the shock front.
Although its temperature is not plotted, He, which carries significant
energy, is included in these calculations and explains why the Mg
temperature levels off before eventually equilibrating with the proton
temperature.\label{fig-tempequil}}
\end{figure}

We expect that the observed X-ray emission will come primarily
from the blast wave, which travels $\approx 3500 \kms$ ($T_s \approx 17
\keV$) into \HII\ region gas with atomic density of $\sim 100 \pcc$
\citep{chev95,lund99}.  If the blast wave has encountered the ring,
with density $\sim 10^4 \pcc$, shocks of velocity 300 -- $1000 \kms$
($T_s \approx 0.1$ -- $1.4 \keV$) would be transmitted (see
\citealt{pun02}).  Figure~\ref{fig-tempequil} depicts ion and electron
temperature equilibration through Coulomb collisions behind these two
representative shocks for two different initial electron temperatures.
In the \HII\ region, having density of $\sim 100 \pcc$, the faster
shock will take $\sim 10^3$ years to reach temperature
equilibration. However, in the slower shock entering the dense ring
the electron and ion temperatures will equilibrate in less than a few
months.
 
In fact, both of these shocks can be approximated fairly well by
constant electron temperature models. Since the first interaction with
the \HII\ region came at day $\approx 1000$, the blast wave should
have an ionization age of $\sim 5 \times 10^{10}$ cm$^{-3}$
s. Therefore, even in the case where no electron heating occurs at the
shock front, most of the plasma ($\sim 90\%$ by volume) has nearly
constant (within a factor of 2) electron temperature.  Because of its
much larger ionization age ($\sim 10^{13}$ cm$^{-3}$ s), a slower
shock would be in a regime where most of its emission comes after
temperature equilibration has occurred. So, in view of the fact that
the {\it pshock} models are much faster to calculate, for most of our
fits we have assumed that constant electron temperature models are
adequate to describe the actual post-shock temperature structure in
these shocks.  We will confirm this assumption for the faster shock in
\S\ref{sec-disc} using the {\it npshock} models.

\section{Analysis} \label{sec-anal}

\subsection{CCD Spectra} \label{sec-analCCD}

The superb spatial resolution of {\it Chandra} ($\approx 0\farcs5$) and the
quality ($\approx 9250$ source counts) of the CCD data allow us to
perform spatially resolved spectroscopy of the remnant.  We analyze
the spectrum of the ($\approx 1\farcs2 \times 1\farcs0$) X-ray remnant
in three stages of increasing spatial detail. First, we consider the
spectrum of the entire remnant.  Then we analyze the spectra of the
eastern and western halves.  Finally, we analyze the spectra from the
six regions defined in \S\ref{sec-obsCCD}.  For each stage of our
analysis we use the {\it XSPEC} package and vary the binning used in
our fits: a minimum of 100 counts/bin for the spectrum of the entire
remnant, 50 counts/bin for the spectra of the remnant's halves, and
16 counts/bin for the six regions.  With these binnings, each spectrum
has a similar number ($\approx 70$) of spectral bins.

As described in \S\ref{sec-physics}, the hydrodynamics of the
interaction of the SN ejecta with the CSM is probably complex.  It
would be naive for us to hope to discriminate all its possible details
by analyzing the X-ray spectra. Nonetheless, it is reasonable to
assume that shock physics is responsible for the emission, so we make
use of various shock models in order to fit the observed X-ray
spectra.  As an approximation to the actual situation, we fit the data
with two-shock models in which the faster shock represents the
interaction of the blast wave with the \HII\ region gas, while the
slower shock represents the interaction with considerably denser gas,
such as that found in the ring.

In order to infer the overall chemical abundances and the
interstellar/circumstellar X-ray absorption column density, we first
analyze the spectrum of the entire remnant, assuming that the
abundances are uniform throughout the remnant. The portion of the
spectrum with good statistics ($E \approx 0.5$ -- $4 \keV$) is
sensitive only to the following abundances: N, O, Ne, Mg, Si, S, and
Fe.  We allow these abundances to vary, while fixing the abundances of
He and C at values (2.57 and 0.09 respectively) found for the
circumstellar ring \citep{lund96}, and Ca and Ni at values (0.34 and
0.62 respectively) representative of the Large Magellanic Cloud (LMC)
\citep{russ92}. Unless otherwise specified, in this paper we present
abundance values with respect to their Solar values
\citep{ande89}. Once we derive the X-ray absorption column density and
chemical abundances from the fit to the spectrum of the entire
remnant, we adopt these values for the analyses of the spectra of the
different regions.

\begin{figure}
\epsscale{.45}
\plotone{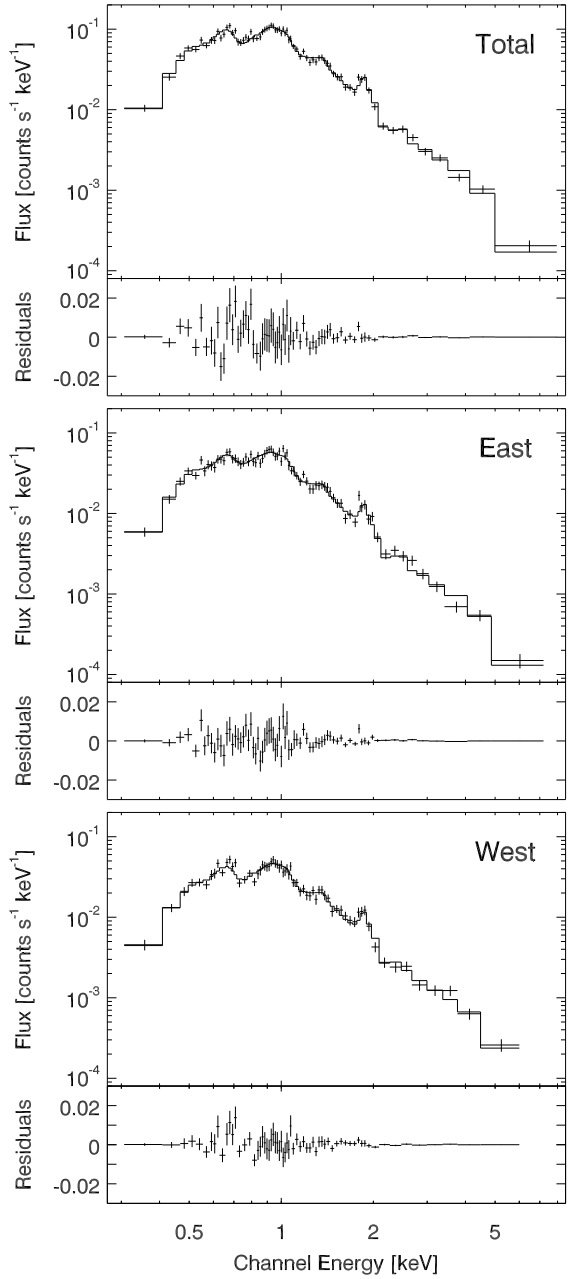}
\caption{CCD Spectra (error bars) with model fits (solid lines) and
residuals. The top panel shows the spectrum of the entire remnant
(binned with a minimum of 100 counts per bin) with the single shock
fit given in Table~\ref{tab-SSfits}.  The bottom panels show the
spectra from the eastern and western halves of the remnant (binned
with a minimum of 50 counts per bin).  The fitted model for these
spectra is one which has the same fast shock component in both halves
but with an additional slow shock component in the
east.\label{fig-CCDfullEW}}
\end{figure}

Table~\ref{tab-SSfits} presents the results from single shock fits to
the spectra of the whole remnant (Figure~\ref{fig-CCDfullEW}) and to
the eastern and western halves.  For these fits, the {\it vpshock}
model was used with the velocity shift of SN 1987A ($v_{shift} = 287
\kms$; \citealt{meab95}). For our fits, we used LMC abundances
\citep{hugh98,russ92} for the absorption since most of the neutral gas
along the line of sight towards SN 1987A is in the LMC \citep{fitz90}.
We find a neutral hydrogen absorbing column of $n_H = 2.5 \times
10^{21} \col$, close to that found from analysis of the damped
Hydrogen \La\ absorption wings of nearby stars \citep{fitz90,scud96}.
If we adopt Solar abundances for the neutral component, our fits
require a smaller value for the column density ($n_H = 1.9 \times
10^{21} \col$), while the other shock parameters remain within
the ranges given in Table~\ref{tab-SSfits}. Correcting for
interstellar absorption and assuming a distance of 50 kpc to the
remnant, we calculate an X-ray luminosity (in the 0.5 -- 2.0 keV band)
of $1.0 \times 10^{35}$ erg s$^{-1}$.  This is $\approx 10\%$ larger
than the luminosity found by \citet{park01} due to the lower value for
$n_H$ that they obtained.  The elemental abundances derived from the
single {\it vpshock} model fit to the X-ray spectrum of the entire
remnant are (with 90\% confidence errors): N = $0.41^{+0.29}_{-0.19}$;
O = $0.07^{+0.03}_{-0.02}$; Ne = $0.12^{+0.06}_{-0.05} $; Mg = $0.11
\pm 0.06$; Si = $0.30^{+0.10}_{-0.09}$; S = $0.38^{+0.30}_{-0.28}$; Fe
= $0.11^{+0.07}_{-0.04}$.

When we split the remnant into two halves and fit the spectra with
single shock models, we find that the western half of SNR 1987A
appears to require slightly hotter plasma than the eastern half.  This
result could imply either that the blast wave in the western half is
faster or that there is an additional component of cooler gas present
in the east.  Due to the presence of the optical spots on the ring, as
the associated soft X-ray spots \citep{park01}, we explore the later
interpretation by considering models which contain two shock
components.

The two-shock models improve the quality of the fits to the X-ray
spectrum of the entire remnant as well as to the eastern half (with
$\chi^2_\nu \approx 1$). The derived absorption column density and
abundances (assumed the same for both shock components) as well as the
parameters of the faster shock (temperature and ionization age) are
almost identical to those derived from the single shock models. The
second shock component has an electron temperature of $\approx 0.4
\keV$ ($v_s \approx 500 \kms$) and an ionization age two orders of
magnitude greater than that of the faster shock. This result is
consistent with our picture where the slower shock is a result of the
blast wave partially interacting with a much denser gas.

The bright optical spots are primarily located on the eastern half of
the remnant.  Perhaps the observed brightness asymmetry in the X-ray
image is related to this fact.  To test whether slow X-ray emitting
shocks associated with the optical spots could be responsible for the
observed X-ray asymmetry we construct the following model.  We assume
the fast shock component has the same parameters in the east and in
the west, while an additional component (a slow shock) is present only
in the eastern half. This model gives a very good fit to the X-ray
spectra of both halves (Figure~\ref{fig-CCDfullEW}), with $\chi^2_\nu
= 1.05$ $(\nu = 70)$ and $0.91$ $(\nu = 63)$ for the eastern and western
halves, respectively.  The blast wave has $T_e = 2.7 \pm 0.2 \keV$ and
$n_e t = (6.5 \pm 0.5) \times 10^{10}$ cm$^{-3}$ s, while the slower
shock in the eastern half has $T_e = 0.4 \pm 0.1 \keV$ and $n_e t \sim
10^{13}$ cm$^{-3}$ s.  We find that the slow shock component accounts
for $7\pm2\%$ of the flux (0.3 -- 6 keV) in the eastern half.  As we
discuss in \S\ref{sec-disc}, at this level, a slow shock component
cannot account for all of the observed flux asymmetry.

\begin{figure}
\epsscale{.8}
\plotone{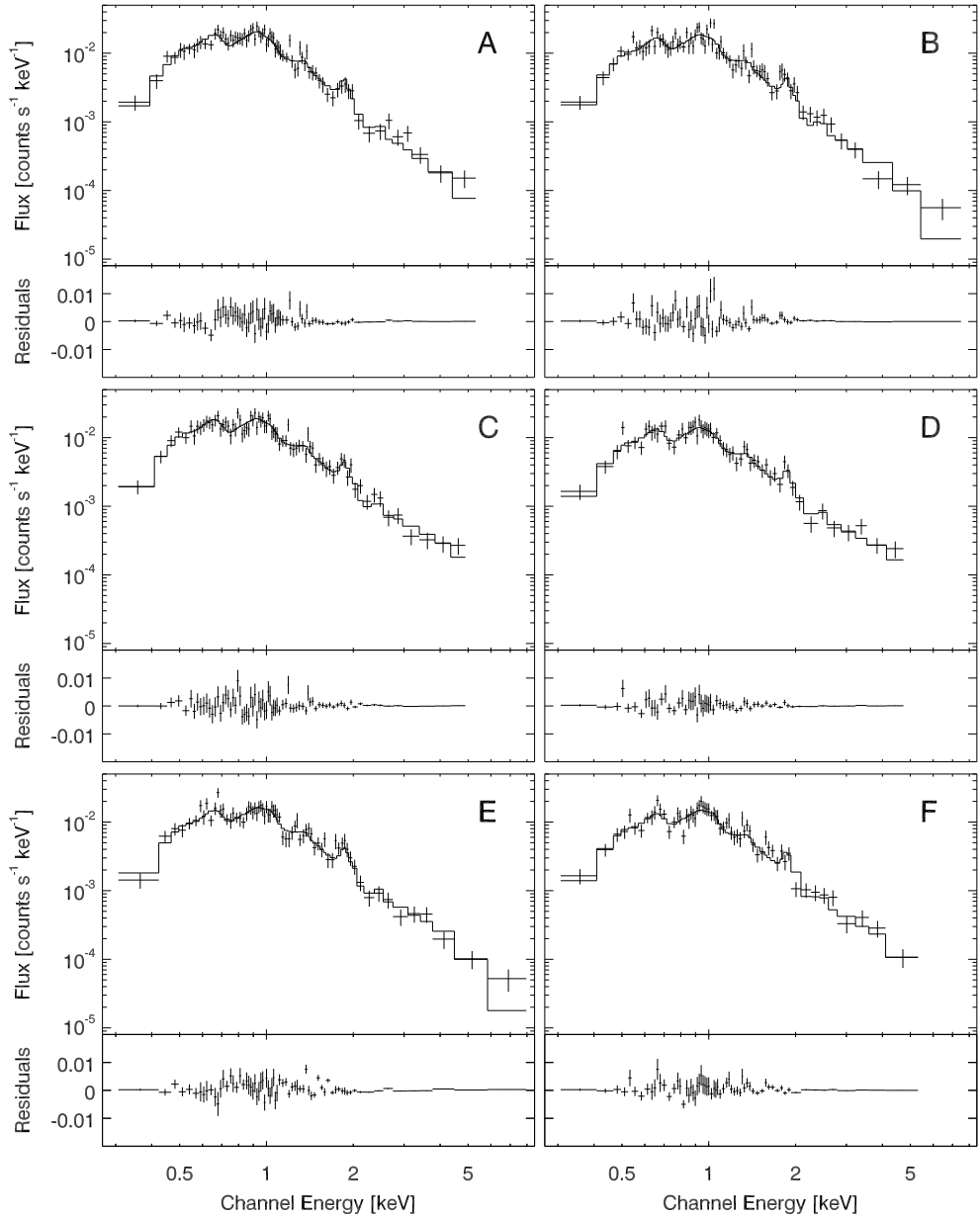}
\caption{The CCD spectra (error bars) with single shock model fits
(solid lines) and residuals for the 6 regions (A -- F) shown in
Figure~\ref{fig-regions}. These spectra are binned with a minimum of
16 counts per bin.\label{fig-CCDreg}}
\end{figure}

Finally, the spectra from the six regions shown in
Figure~\ref{fig-regions} allow us to study the variations of the shock
parameters on a smaller scale. To do so, we fit a single shock model
to the X-ray spectrum of each of these regions
(Figure~\ref{fig-CCDreg}).  The limited photon statistics at this
spatial resolution do not permit us to make significant inferences
from two-shock models. All of these six `local' shocks share the same
abundances and X-ray absorption (both derived from the entire remnant
spectrum). The results of these fits indicate that, while the
ionization times are very similar ($n_e t \approx 6 \times 10^{10}$
cm$^{-3}$ s), the inferred post-shock temperatures vary around the
remnant (see Table~\ref{tab-regions}).  The general behavior of this
variation confirms that the western half of the remnant is hotter than
the eastern half. We note that region A, which contains Spot 1, the
brightest optical spot, has a significantly lower temperature than the
rest of the remnant.  This is a clear indication in the X-ray band
that the blast wave is interacting with denser gas at that location.
This behavior is not seen in the spectrum of region C, however, which
also contains several bright optical spots. This may be due to the
presence of bright hard X-ray emission from the blast wave in region C
\citep{park01} which could overwhelm the softer emission produced by
the less developed optical spots. Alternatively, this may indicate
that the spots in region C are reacting differently in the X-ray than
Spot 1.

\subsection{Dispersed Spectrum} \label{sec-analdisp}

Individual emission lines are hard to identify in the undispersed CCD
spectrum due to its low spectral resolution.  The dispersed spectrum,
however, allows us to identify and measure the fluxes in the brightest
lines independently of any model.  In principle, the ratios of the
fluxes of various individual lines offer powerful diagnostics of the
emitting plasma.  Unfortunately, the low count rate in the dispersed
spectrum limits our ability to exploit this advantage. We can only use
the observed line fluxes to confirm our analysis above, showing that
the observed line strengths are indeed consistent with those predicted
by shock models.

From the dispersed spectrum we extract the fluxes of 9 lines labeled
in Figure~\ref{fig-dispspec}.  Most of these lines are clearly
apparent in the spectrum, while a couple of the lines (the Si lines)
are predicted by the shock models to be bright.  Measurements of upper
limits to the fluxes in these lines therefore provide additional
constraints to the shock models.  No other bright lines ($> 8$ counts)
are identified in the 0.3 -- $6.0 \keV$ energy range.

Given the poor counting statistics of the lines, we cannot accurately
measure their fluxes by fitting them with Gaussian profiles. Instead,
we integrate the fluxes within a given velocity range of line center.
As we discuss in \S\ref{sec-lineprof}, most of the emission in a line
is contained within $\pm 5000 \kms$ of line center.  Therefore, these
limits are used to extract the fluxes in the \La\ lines. With the
present spectrum we are unable to isolate the individual components of
the \Ka\ triplet lines.  For the shock parameter range established
above we suspect small, but perhaps significant, contributions from
the forbidden components of the triplets.  Therefore, for the \Ka\
lines we integrate the flux between $-5000 \kms$ and $ v_{forb} + 5000
\kms$ from the resonance line, where $v_{forb}$ is the velocity offset
of the forbidden component from the resonance line (4000 -- $7000
\kms$ for the observed lines).  From the fluxes extracted from these
ranges, we subtract a background `continuum' level (either
real continuum or underlying emission from other faint lines)
determined from regions extending $7500 \kms$ beyond the extraction
velocity range.  Our results are shown in Table~\ref{tab-linefluxes}. The
tabulated errors include both statistical and systematic errors in the
flux measurements. We estimate systematic errors introduced in the
`continuum' subtraction by unidentified lines by simulating our
technique on model spectra.  We find 7 lines with S/N $\approx 3$, but
only upper limits to the fluxes of \OKa\ and \SiLa.

\begin{figure}
\epsscale{.5}
\plotone{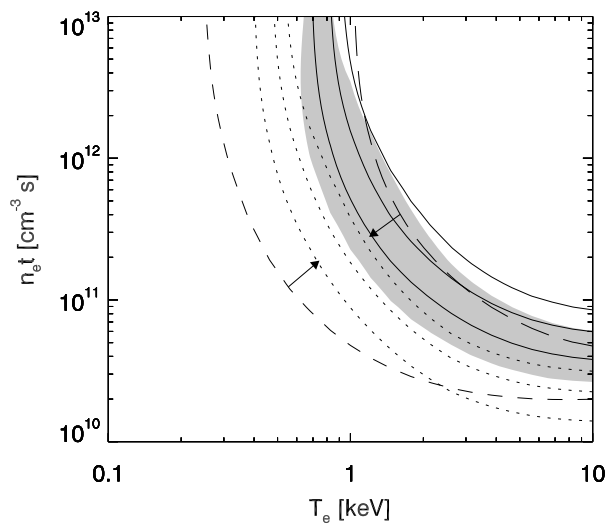}
\caption{Lines show the regions of $(T_e,n_e t)$ phase space that are
permitted by the observed ratios of \La\ to \Ka\ line fluxes. For Mg
(solid) and Ne (dotted) the middle line shows the observed ratio,
while the two other curves bracket the $1\sigma$ errors.  For O
(dashes) a lower limit to the ratio is shown, and for Si (long dashes)
an upper limit is plotted.  The regions of phase space allowed by
these limits are shown by the arrows which connect to the curves.  The
shaded region shows the $90\%$ confidence region for $T_e$ and $n_e t$
obtained by fitting a shock model to all the observed line
fluxes.\label{fig-LK}}
\end{figure}

Since we observe both \La\ and \Ka\ lines for O, Ne, Mg, and Si, we
can test whether the shock models can accurately reproduce their flux
ratios.  Since abundances are always in question when fitting spectra,
these abundance-insensitive line ratios offer a test of the underlying
shock emission models.  We use the {\it XSPEC pshock} model to
calculate the theoretical line fluxes, assuming interstellar
absorption in the LMC with $n_H = 2.5 \times 10^{21} \col$.
Figure~\ref{fig-LK} shows the regions of $(T_e,n_e t)$ phase space
that are permitted by the observed ratio of \La\ to \Ka\ flux for each
element.  All of the observed ratios define a similar region of phase
space.  While the Ne and Mg regions are confined by the limits set by
O and Si, they do not overlap with each other within $1\sigma$ errors
(however, they are consistent at the $2\sigma$ level).  This
discrepancy is not too surprising since there is clearly a
distribution of shocks present.  The ions that form the Mg lines are
formed at higher temperatures than those that produce the observed Ne
emission, and therefore can represent emission from hotter plasma.

The best fit model to all the line strengths has a $\chi^2_\nu = 1.28$
$(\nu = 2)$.  Since the data set is limited to 9 lines, the model is
poorly constrained when we fit for the abundances (N, O, Ne, Mg, and
Si) as well as temperature and ionization age.  The shaded region in
Figure~\ref{fig-LK} shows the 90\% confidence region for $T_e$ and
$n_e t$.  The parameters found in the CCD analysis fall within this
confidence region --- that shock model can reproduce the line fluxes
seen in the dispersed spectrum.  We find abundances, though not well
constrained, which are consistent with those determined through
analysis of the CCD spectrum.  In particular, the presence of the
bright \NLa\ line requires that the gas be nitrogen enriched (N/O
$\gtrsim 0.7$ by number).

If we fit models to the entire dispersed spectrum , we can constrain
the shock model more tightly due to the added information provided by
the continuum and the absence of higher energy lines in the spectrum.
Since the MEG and HEG spectra have different responses, we fit them
simultaneously within {\it XSPEC}. In order to obtain an acceptable
S/N per bin, we re-bin the spectra so that each bin has at least 10
counts (yielding 57 and 23 bins for the MEG and HEG spectra
respectively).  This procedure degrades the fine resolution of the
grating observations somewhat, but the resolution is still good
($\approx 2000 \kms$) where the lines are brightest.

Since we expect a blast wave velocity of $\approx 4000 \kms$ we expect
similar widths in the line profiles (see
\S\ref{sec-lineprof}). Therefore we smooth the shock model with a
Gaussian profile ({\it gsmooth}) with the velocity width a free
parameter in the fit.  Using the abundances and absorbing column
determined in \S\ref{sec-analCCD} we fit the spectra using the {\it
vpshock} model with the velocity shift of SN 1987A.  The best fit has
$\chi^2_\nu = 1.01$ $(\nu = 76)$ with $T_e = 2.9 \pm 0.4 \keV$ and $n_e
t = (5 \pm 1) \times 10^{10}$ cm$^{-3}$ s, similar parameters to those
found from the CCD spectrum.  The fact that the grating spectrum was
taken $\approx 1$ year earlier than the CCD spectra ($\approx 10\%$ of
the shock age) might explain why a marginally lower ionization age is
required for this spectrum.  The lines have full width at half maximum
(FWHM) of $3300 \pm 3000 \kms$.  Although not required by the data, a
two-shock model fit, with $\chi^2_\nu = 1.06$ $(\nu = 72)$, has an
additional soft component ($T_e \sim 0.3 \keV$) which contributes
$<10\%$ of the total flux.

\subsection{Line Profile} \label{sec-lineprof}

\begin{figure}
\epsscale{.5}
\plotone{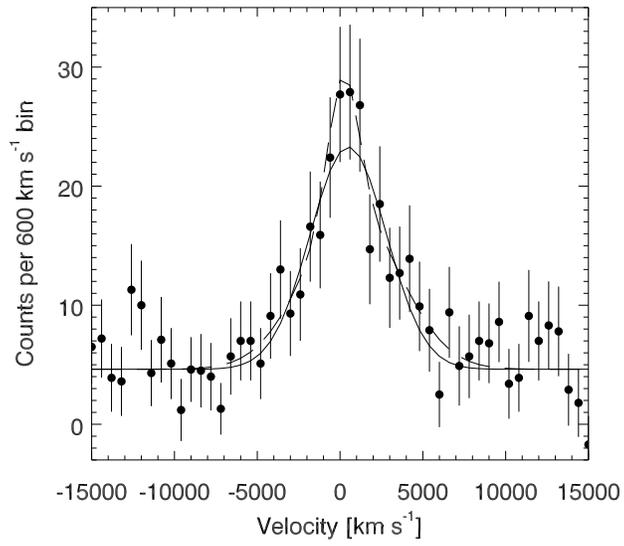}
\caption{The combined X-ray line profile binned at $600 \kms$.  Also shown
are the single (solid) and double (dashed) Gaussian
fits. \label{fig-lineprof}}
\end{figure}

The high resolution grating spectrum provides us access to the X-ray
line profiles at a detector resolution of $\approx 700 \kms$.
Unfortunately, the individual lines observed each have too few counts
to make reasonable tests of their profiles.  Therefore we create a
combined line profile by adding together several of the brightest
lines.  For this analysis we include the \NLa, \OLa, \NeKa,
\NeLa, \MgKa, \MgLa, and \SiKa\ lines.  The combined line profile is
shown in Figure \ref{fig-lineprof}. This profile has decent statistics
--- it contains $\approx 180$ counts spread over $\sim 10,000 \kms$, so
we can obtain an average S/N per bin of $\sim 3$ with $600 \kms$ bins.

It is acceptable to include the \Ka\ triplet lines in the combined
profile since, for the parameter range we have established for the
emitting plasma, we expect the contribution to the profile from the
forbidden components of these lines to be small ($\lesssim 7\%$).  As
a confirmation, there is no evidence of a bulge in the profile at 4000
-- 7000 $\kms$, where the contributions from the forbidden components
would be largest.  We calculate that the contamination of the profile by
other lines falling within $\approx \pm 5,000 \kms$ of the included
lines (mostly Fe emission near the Ne lines) is small ($\lesssim
15\%$) for the typical shock conditions present.

A glance at the profile shows it is centrally concentrated with FWHM
of $\approx 5000 \kms$ and centroid at $\approx 500 \kms$ (note that
SN 1987A has a relative velocity of $287 \kms$). We fit the profile
with functions consisting of a constant background plus one or two
Gaussian components ($e^{-(v - v_{shift})^2 / 2\Delta v^2}$).  We have
performed these fits with various bin sizes (300 -- 1000 $\kms$) to
test the stability of our results.  Since the width of the profile may
be comparable the size of the velocity bins, the Gaussian profiles are
properly integrated over the bin limits. We find that the $\chi^2_\nu$
statistic remains acceptable for these fits but its value wanders as a
function of bin size ($\chi^2_\nu \approx .8$ -- $1$).  The fitted
parameters remain steady within the limits of their errors.  Neither
the single nor double Gaussian fits are statistically superior, we
have included the double Gaussian fit to test for the presence of a
narrow component to the line profile. The resulting fits are shown in
Figure~\ref{fig-lineprof} and the fit parameters are tabulated in
Table~\ref{tab-linefits}.

The fitted velocity widths, $\Delta v$, are degraded due to the finite
image size of the remnant. The remnant's emission extends nearly
$1\farcs5$ along the dispersion direction.  The combined profile is
dominated by Ne and Mg emission lines near $1.2 \keV$.  At the
dispersion of the MEG the spatial extent of the remnant accounts for
$\approx 900 \kms$ at $1.2 \keV$.  For the broad Gaussian component
this effect is small enough to ignore. If a narrow component is
present, it would be unresolved.  Though poorly constrained, its
possible presence supports the idea that slow shocks ($\lesssim 1000
\kms$) associated with the optical spots may be responsible for some
of the X-ray emission from the remnant.

There are two astrophysical broadening mechanisms acting on the line
profile: thermal broadening and broadening due to bulk fluid motions.
Since both of these motions are induced by the passage of the blast
wave, their magnitudes are dependent on its velocity.  Therefore we
may use the measured velocity widths to constrain the speed of the
blast wave.

Assuming that the ions have Maxwellian velocity distributions, the
thermal widths of the emission lines are given by $\Delta v_{therm,i}
= \sqrt{kT_i/m_i}$, where $T_i$ is the ion temperature and $m_i$ is
the mass of the ion.  The ion temperatures depend on the degree of
equilibration occurring behind the shock front, by either collisional
or collisionless processes.  In the case of no temperature
equilibration, the thermal widths would be independent of ion mass and
would be given by $\Delta v_{therm} = \sqrt{3/16} v_s$.  However, if
temperature equilibration has occurred, the widths would be reduced by
a factor as large as $\sqrt{\bar{m}/m_i}$.

\begin{figure}
\epsscale{1.}
\plotone{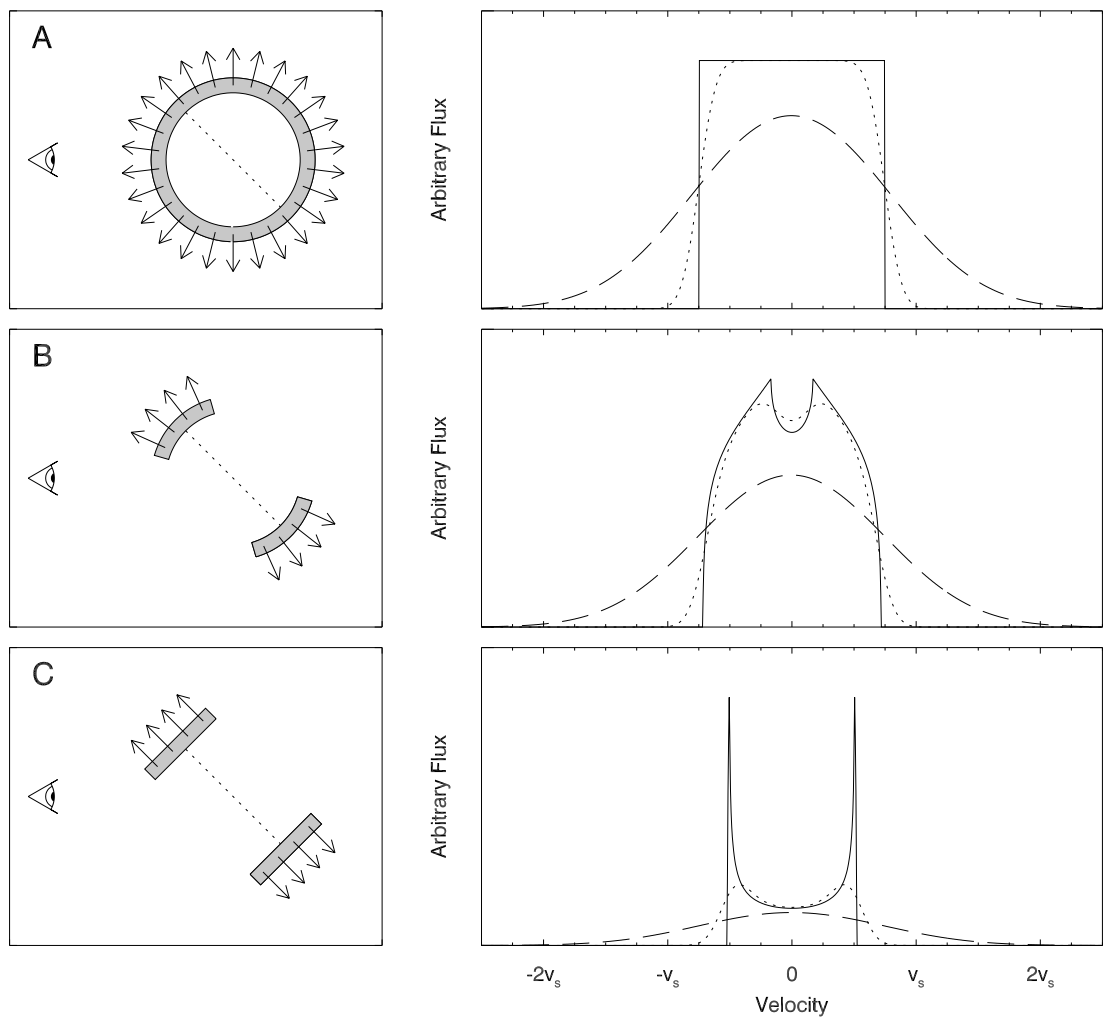}
\caption{The panels on the left show three possible geometries and
flow fields for the X-ray emitting plasma: (A) an expanding spherical
shell of gas, (B) a radially expanding equatorial shell of gas, and
(C) a cylindrical shell expanding parallel to the equatorial
plane. The dotted lines show the equatorial plane which is inclined at
$\approx 45^\circ$ from the line of sight.  The geometries are assumed
to have cylindrical symmetry. The panels on the right show the line
profiles produced by these geometries. Solid lines show the unsmoothed
bulk motion profiles, the dotted lines show the bulk profiles smoothed
with thermal widths representing the case where ion temperature
equilibration is complete, and the dashed lines show the bulk profiles
smoothed assuming that no ion temperature equilibration has
occurred. \label{fig-simprof}}
\end{figure}

Since the line profile contains emission from the entire remnant, we
observe emission from fluid elements traveling in many directions with
bulk speeds comparable to the blast wave's speed.  The precise
distribution of emission as a function of line-of-sight Doppler
velocity depends on the hydrodynamics and kinetics of the blast wave's
interaction with the surrounding material.  However, we can create
simplified models for the bulk motion profile based on geometric
models for the blast wave.  Figure~\ref{fig-simprof} shows the three
models we consider: emission from a spherical shell of gas expanding
radially with constant velocity, emission confined to the equatorial
region of an expanding spherical shell, and a cylindrical shell
expanding with constant velocity parallel to the equatorial plane.
The equatorial geometries are chosen to mimic the observed geometry of
the reverse shock, which has a latitudinal extent of
$\pm30^{\circ}$. The line profiles that result from these geometries
are also shown in Figure~\ref{fig-simprof}. For the equilibrated thermal
widths, we have assumed $m_i = 22 m_p$, representative of Ne and Mg. In
the limit of no temperature equilibration the thermal widths are
comparable to the bulk widths and the resulting line profiles more
closely resemble Gaussians.

For the profiles smoothed with the unequilibrated thermal width we
find that $\Delta v = 0.77 v_s$ for the spherical model and $\Delta v
= 0.74 v_s$ for the equatorial models. If thermal equilibration is
complete then the simulated profiles are narrower, with $\Delta v =
0.63 v_s$ and $\Delta v \approx 0.5 v_s$ for the spherical and
equatorial models respectively.  Using these relations we calculate
from the measured widths given in Table~\ref{tab-linefits} that $v_s$
falls in the range 2500 -- $6500 \kms$.  However, for the ionization
age determined from fits to the dispersed spectrum ($n_e t \approx 5
\times 10^{10}$ cm$^{-3}$ s) the shocked plasma should not have had
time to substantially equilibrate its ion temperatures (see
Figure~\ref{fig-tempequil}).  This idea is supported by the fact that
the profile is well fit by a Gaussian, and does not show structure
like that seen in the equilibrated model profiles. Moreover, we favor
the equatorial geometries for the blast wave emission. Using these
model parameters, and a velocity width representative of the broad
Gaussian component ($\Delta v = 2500 \pm 500 \kms$), we infer that $v_s
= 3400 \pm 700 \kms$.  This velocity is consistent with those obtained
by measuring the expansion of the radio and X-ray images and
that predicted by hydrodynamic theory.

\section{Discussion} \label{sec-disc}

The blast wave velocity implied by the width of the X-ray line profile
indicates that the shock temperature should be $\approx 17 \keV$.
This is much higher than the electron temperature measured in our fits
to the X-ray spectra ($T_e \approx 2.5 \keV$).  Though not surprising
given the discussion in \S\ref{sec-physics}, this result is direct
observational evidence of the existence of unequal electron and ion
temperatures behind a fast shock ($\beta = T_{e0} / T_s < 1$).  Up to
now we have used models in which the post-shock electron temperature is
constant. In order to confirm the validity of this assumption, and to
constrain the amount of electron heating at the shock front itself, we
make use of the {\it npshock} models which account for the
equilibration of electron and ion temperatures through Coulomb
collisions.  We fit the CCD spectrum of the entire remnant with the
abundances and interstellar absorption given in \S\ref{sec-analCCD}.

Since line excitation is caused primarily by collisions with
electrons, the X-ray spectrum is insensitive to the ion temperatures,
and thus to the shock speed. Therefore the parameter $T_s$ (and thus
$\beta$) is poorly constrained by our spectral analysis.  With this in
mind, we perform various fits each having a fixed value of $\beta$,
fitting for the initial electron temperature $T_{e0}$ and ionization
age.  Fits with $\beta = 0.05, 0.1$, and $0.2$ give results of
comparable statistical significance ($\chi^2_\nu$ of 1.13, 1.05, and
1.07, with $\nu = 69$) with initial electron temperatures of $1.6,
1.7$, and $2.0 \keV$ respectively, and ionization ages of $\approx 5
\times 10^{10}$ cm$^{-3}$ s.  Note that all of the fitted initial
electron temperatures are lower than the value ($2.5 \keV$) found in
the constant electron temperature fits.  This is true because the
temperature determined in the {\it pshock} model fit represents the
mean electron temperature of the plasma.  An {\it npshock} model fit
with a lower value for $\beta$ has a higher shock temperature, thus a
faster heating rate, and so requires a lower initial electron
temperature in order to produce a plasma with the same mean electron
temperature. That all of the $\beta$ model fits have similar
$\chi^2_\nu$ values indicates that the spectrum is most sensitive to
the mean electron temperature and less so to the actual temperature
distribution.  This is primarily due to the fact that the electron
temperature remains nearly constant over the volume behind the shock
(it varies by at most a factor of $\approx 2$ in the $\beta = 0.05$
model).  This result supports our use of the constant electron
temperature models as representative models in our analysis in
\S\ref{sec-anal}.

The CCD spectrum alone cannot constrain the value of $\beta$ since it
is insensitive to the shock temperature.  However, we note that the
$\beta = 0.1$ model seems the most realistic since its shock
temperature, $T_s = 17.3 \keV$, is very close to what is expected from
a shock with velocity of $\approx 3500 \kms$.  Indeed we can use the
shock velocity results of \S\ref{sec-lineprof} and fix the shock
temperature at $16.5 \keV$ in order to constrain $\beta$.  For this
fit we find a good fit, $\chi^2_\nu = 1.04$ $(\nu = 70)$ with $T_{e0} =
1.8^{+0.4}_{-0.1} \keV$ or $\beta = 0.11^{+0.02}_{-0.01}$.  It is
worth mentioning that models where $T_{e0}$ is fixed at a very low
value (as would be the case if no collisionless electron heating
occurred at the shock front) are not consistent with the data. This
amount of equilibration for a $\approx 3500 \kms$ shock is consistent
with the results of \citet{ghav01}, who found that $\beta$ decreases
with increasing shock strength.  However, the observed electron
temperature is not low enough to imply that particle acceleration has
taken a significant amount of energy from the plasma, as found in the
blast wave of 1E 0102.2-7219 \citep{hugh00}.

The X-ray image (as well as images of the radio source) show a
noticeable brightness asymmetry between the eastern and western halves
of the remnant.  Our analysis in \S\ref{sec-analCCD} quantified the
X-ray brightness asymmetry as the eastern half having 15\% more flux
(in the 0.3 -- 6 keV band) than the western half.  However, this
result is based on our choice of the image centroid as the center of
the X-ray image. Clearly the level of flux asymmetry depends on this
choice.  To explore the possible extent of the asymmetry, we performed
a similar spectral analysis with the center defined from an ellipse
which was fit to the shape of the X-ray image (see
Figure~\ref{fig-regions}).  In this case the asymmetry between the
X-ray fluxes of the eastern and the western halves increases to
$\approx 50\%$. We fit one- and two-shock models to the spectra in a
similar manner as described in \S\ref{sec-analCCD} and find similar
physical parameters at about the same statistical
significance. Unfortunately, absolute X-ray astrometry was not
possible with our data so we are unable to precisely constrain the
level of asymmetry any better than 15 -- 50\%.

The presence of slower X-ray emitting shocks where the blast wave has
struck the inner ring (representing $<5\%$ of the total X-ray flux)
cannot account for the observed level of flux asymmetry.  Therefore,
this asymmetry must be due primarily to a global asymmetry in the
interaction of the SN blast wave with the circumstellar environment.
The blast wave must be interacting with denser material in the east.
The luminosity of the shocked plasma is approximately proportional to
$n_0^2T_s^{-0.6}V$, where $n_0$ is the pre-shock density, $V (\propto
v_s)$ is the volume of shocked gas, and $T_s (\propto v_s^2)$ is the
shock temperature.  Since $v_s \propto n_0^{-1/2}$, the luminosity
goes as $n_0^{2.1}$.  Whatever level of brightness asymmetry that
exists is a good estimate of the differences in pre-shock density that
the blast wave is encountering.

The blast wave could be encountering denser material in the east
either because the CSM is asymmetric or because the SN explosion is
asymmetric. If the SN explosion (and therefore the expanding debris)
was symmetric then the CSM must have greater density toward the east
then toward the west. In this case the blast wave would propagate more
slowly toward the east than toward the west. If, on the other hand,
the CSM is symmetric, the blast wave must have propagated further
toward the east to encounter the denser gas closer to the equatorial
ring.  For this to occur, the driving pressure of the SN ejecta must
be greater in the east than in the west, indicating an asymmetric SN
explosion.  It seems reasonable that the SN occurred at the center of
the ring, in which case the appearance of the optical spots primarily
in the east favors the latter hypothesis.  Unfortunately, we cannot
confirm this hypothesis with the present X-ray data, which lacks
sufficiently accurate astrometric registration with respect to the SN
or the optical ring. This is not a trivial task due to the scarcity of
X-ray point sources in the vicinity of SN 1987A.  However, accurate
X-ray astrometry would yield powerful scientific rewards.  Short of
this, the location of the reverse shock is a good proxy for the
location of the blast wave and may provide the best constraints on the
cause of the observed asymmetry (Michael et al. 2002, in
preparation).

Many young SNR have been observed to contain nonthermal X-ray emission
(e.g., \citealt{alle99}).  From an analysis of {\it XMM-Newton} data
\citet{asch01} have suggested that the spectrum of SNR 1987A includes
such a component. We explore this possibility with the {\it Chandra}
CCD spectrum by considering a composite model consisting of a single
shock and a power-law component.  Our best fit, with $\chi^2_\nu =
1.06$ $(\nu = 60)$, is not statistically superior (based on the F-test)
to our previous models.  The photon index of the power-law emission is
$\Gamma = 2.2\pm0.3$. The shock has $T_e = 1.4 \pm 0.2 \keV$, $n_et
\approx 10^{11}$ cm$^{-3}$ s, and has abundances which are greater by
factors of $\approx 2$ -- 3 compared to the values derived
previously. The observed X-ray flux/luminosity of the power-law
component is about 57\% of the total in the 0.3 -- 6 keV range and
$\sim 75\%$ in the 1.4 -- 8 keV range.  The upper limit to a possible
point source of X-ray emission in SNR 1987A is $\approx 13\%$ in the 2
-- 10 keV band \citep{park01}. Therefore, an embedded pulsar or its
wind nebula cannot be responsible for such a nonthermal component.  On
the other hand, the fast shocks themselves might produce X-ray
synchrotron radiation if they could accelerate electrons to
sufficiently high energies.  In the case of the most efficient
diffusive shock acceleration of particles (e.g., \citealt{joki87}),
when the shock front is parallel to the magnetic field, we estimate
that a magnetic field of strength $\gtrsim 3 \times 10^{-4}$ G must be
present in order for the shocks to accelerate electrons to
sufficiently high energies within the age of the remnant.

With the success of the shock models at fitting the {\it Chandra} spectra,
we find that a power-law component is not required by the data.
Moreover, the abundances required for such a fit seem unreasonably
high.  However, the present spectra do not contain significant flux
above a few keV and therefore the presence of a hard power-law tail can
not be ruled out.  

The possible presence of a narrow component to the X-ray line profile
is intriguing although at present this result is not very
statistically significant.  Confirmation of a narrow component would
support the existence of X-ray emission from lower velocity shocks
entering the ring.  As the remnant gets brighter and the blast wave
penetrates more of the ring, the X-rays produced by the slower shocks
will contribute an increasing percentage of the total X-ray luminosity
\citep{bork97b}. One manifestation of this will be a growing narrow
component in the line profile.  Moreover, since the efficiency of a
shock at producing a given line depends on its velocity, different
lines are likely to have significantly different profiles. For example
the \SiKa\ and \SiLa\ lines are formed only in the faster shocks while
the \OLa\ line will have a significant contribution from slower
shocks. Thus we would expect the Si lines to appear broader. Detection
of differences in the line profiles will help resolve the different
shock components present in the remnant.

The overall spectrum will also change as slower shocks begin to
dominate, adding a softer emission component. The structure of SNR
1987A and its emission is changing significantly on a time scale of
less than a year, therefore much can be learned about the development
of the remnant and its shocks through continued observations of the
remnant in the X-ray band.

\section{Summary} \label{sec-summary}

In conclusion, we summarize our findings:

\begin{itemize}

\item{The X-ray spectrum (including the resolved line fluxes) is well
represented by an adiabatic, plane parallel shock model with mean
electron temperature of $\approx 2.6 \keV$ and ionization age of
$\approx 6 \times 10^{10}$ cm$^{-3}$ s.  We find elemental abundances
that are subsolar ($\approx 0.2$), with an enhanced nitrogen
abundance (N/O $\approx 0.8$ by number).}

\item{There are indications that a second shock component (with $T_e
\approx 0.4 \keV$ and $n_e t \sim 10^{13}$ cm$^{-3}$ s) is present in
the eastern half of the remnant.  This slower shock component (with
$v_s \approx 500 \kms$) may be related to the shocks ($v_s \lesssim
250 \kms$) that produce the optical spots on the equatorial
ring. Currently, the slow shock can account for only $\approx 4\%$ of
the total X-ray emission from the remnant.}

\item{By investigating simulated profiles based on simple models for
the blast wave's geometry, we infer from the observed X-ray line
profile that the blast wave has a velocity of $3400 \pm 700 \kms$.
Future observations of X-ray line profiles should show different
widths for different lines and increasing flux in a narrow component.}

\item{The measured electron temperature in the spectra and the width
of the line profile provide direct evidence for incomplete
electron-ion temperature equilibration behind the blast wave.
Assuming the shock temperature indicated by the line profile, we
measure the level of collisionless electron heating at the shock front
at $T_{e0} / T_s = 0.11^{+0.02}_{-0.01}$.}

\item{We constrain the level of east-west brightness asymmetry at 15
-- 50\%. This asymmetry is caused by the blast wave interacting with
denser material in the east.  A better constraint will only be
possible if a more precise astrometric registration of the location of
the SN is undertaken for the X-ray image.  Such a registration will
also help determine if the source of the asymmetry is an asymmetric
circumstellar environment or an asymmetric SN explosion.}

\item{Continued observations in the X-ray band of this rapidly
brightening and changing remnant will help us better understand SN
1987A, its circumstellar environment, and their interaction.}
 
\end{itemize}

\acknowledgments We would like to thank Kazik Borkowski for his help
and expertise with the shock models we used for this analysis. 
This work was supported by NASA through grants NAG5-3313 and NTG5-80.

\clearpage


\begin{deluxetable}{lccc}
\tablecolumns{4}
\tablewidth{0pt} 
\tablecaption{Counts in Dispersed Spectra\label{tab-dispfluxes}}
\tablehead{
\colhead{Spectrum} & \colhead{Region} & \colhead{Background} & \colhead{Source}}
\startdata
MEG $\pm 1^{st}$ order & 785 & 183.1 & 601.9 \\
HEG $\pm 1^{st}$ order & 412 & 157.1 & 254.9 \\
\tableline
Total	               & 1197 & 340.2 & 856.8 \\
\enddata
\end{deluxetable}
\clearpage

\begin{deluxetable}{lccc}
\tablecolumns{4}
\tablewidth{0pt} 
\tablecaption{Single Shock Model Fit Results\tablenotemark{a}\label{tab-SSfits}}
\tablehead{
\colhead{Parameter} & \colhead{Entire Remnant} &
\colhead{Eastern Half} & \colhead{Western Half}}
\startdata
$\chi^2_\nu$  		& 1.17 			& 1.11 			& 0.88 \\
$\nu$			& 62			& 73			& 63  \\
$n_H$ [10$^{21} \col$]  	& $2.5^{+0.4}_{-0.3}$  	& fixed   		& fixed   \\
$T_e$ [keV]  			& $2.6^{+0.4}_{-0.3}$  	& $2.4 \pm 0.2$		& $2.7 \pm 0.3$ \\
$n_e t$ [10$^{10}$ cm$^{-3}$ s] & $5.8^{+2.1}_{-1.5}$  	& $6.0^{+0.9}_{-0.7}$  	& $6.2^{+1.0}_{-0.8}$ \\
$n^2V$ [10$^{57}$ cm$^{-3}$]  	& $6.6^{+0.9}_{-0.6}$ 	& $3.6 \pm 0.3$		& $3.0 \pm 0.2$  \\
$F_X$ [10$^{-13}$ ergs cm$^{-2}$ s$^{-1}$]\tablenotemark{b} 
				& $3.55 \pm 0.06$ 	& $1.88 \pm 0.05$	& $1.62 \pm 0.03$ \\
\enddata
\tablenotetext{a}{With 90\% confidence errors.}
\tablenotetext{b}{The X-ray fluxes are tabulated for the 0.3 -- 6 keV
range.}
\end{deluxetable}
\clearpage

\begin{deluxetable}{lcccccc}
\tablecolumns{7}
\tablewidth{0pt}
\tablecaption{Post-shock Temperature Variation around the Remnant\label{tab-regions}}
\tablehead{
\colhead{Region:} & \colhead{A} & \colhead{B} & \colhead{C} &
\colhead{D} & \colhead{E} & \colhead{F} }
\startdata
$T_e$ [keV] & $1.9\pm0.1$ & $2.3\pm0.2$ & $3.0\pm0.2$ & $3.0\pm0.3$ &
$2.9\pm0.2$ & $2.4\pm0.2$ \\
\enddata
\end{deluxetable}
\clearpage

\begin{deluxetable}{lcccc}
\tablecolumns{5}
\tablewidth{0pt} 
\tablecaption{Line Fluxes Extracted from Dispersed
Spectrum\label{tab-linefluxes}}
\tablehead{
\colhead{Line} & \colhead{Energy} & \colhead{Flux} 
 & \colhead{Flux} & \colhead{S/N} \\
\colhead{} & \colhead{[keV]} & \colhead{[counts]} 
 & \colhead{[$10^{-15}$ ergs s$^{-1}$ cm$^{-2}$]} & \colhead{} }
\startdata
\NLa   	& 0.500	& 10.4 $\pm$ 4.9  & 11.8 $\pm$ 5.6	& 2.1	\\
\OKa	& 0.574 & $<$6.7	  & $<$9.1	   	& ---	\\
\OLa	& 0.654 & 19.1 $\pm$ 6.4  & 15.2 $\pm$ 5.1	& 3.0	\\
\NeKa	& 0.921 & 32.4 $\pm$ 10.4 & 9.7 $\pm$ 3.1	& 3.1	\\
\NeLa	& 1.022 & 30.7 $\pm$ 9.5  & 6.7 $\pm$ 2.1	& 3.2	\\
\MgKa	& 1.352	& 47.1 $\pm$ 12.4 & 5.7 $\pm$ 1.5	& 3.8	\\
\MgLa	& 1.473 & 28.9 $\pm$ 8.9  & 3.4 $\pm$ 1.1	& 3.2	\\
\SiKa	& 1.865	& 29.5 $\pm$ 10.5 & 4.6 $\pm$ 1.6	& 2.8	\\
\SiLa	& 2.007	& $<$10.2 	  & $<$1.7  		& ---	\\
\enddata
\end{deluxetable}
\clearpage

\begin{deluxetable}{lcccc}
\tablecolumns{4}
\tablewidth{0pt} 
\tablecaption{X-ray Line Profile Gaussian Fit Parameters\label{tab-linefits}}
\tablehead{
\colhead{Fit} & \colhead{Flux} & \colhead{$v_{shift}$} 
	& \colhead{$\Delta v$} \\
\colhead{} & \colhead{[counts]} & \colhead{[$\kms$]} 
	& \colhead{[$\kms$]}}
\startdata
Single Gaussian & $180 \pm 20$	& $500 \pm 300$	& $2300 \pm 300$ \\
\tableline
Double Gaussian & $160 \pm 35$	& $700 \pm 400$	& $2900 \pm 700$ \\
                & $45 \pm 35$	& $200 \pm 400$	& $700 \pm 500$	\\
\enddata
\end{deluxetable}
\clearpage
\clearpage


\clearpage

\clearpage

\clearpage

\clearpage

\clearpage

\clearpage

\clearpage

\clearpage

\end{document}